\documentstyle[preprint,aps,prd]{revtex}
\title{Quantum states with negative energy density \\in the
Dirac field and quantum inequalities}
\author{ Hongwei Yu  and Weixing Shu  }
\address
{ Department of Physics and Institute of  Physics,\\ Hunan Normal
University, Changsha, Hunan 410081, China. } \tightenlines
\begin{document}
\maketitle
\begin{abstract}
Energy densities of the quantum states that are superposition of
two multi-electron-positron states are examined. It is shown that
the energy densities can be negative  when  two multi-particle
states have the same number of electrons and positrons or when one
state has one more electron-positron pair than the other. In the
cases in which negative energy could arise, we find that the
energy is that of a positive constant plus a propagating part
which oscillates between positive and negative, and the energy can
dip to negative at some places for a certain period of time  if
the quantum states are properly manipulated. It is demonstrated
that the negative energy densities satisfy the quantum inequality.
Our results also reveal that for a given particle content, the
detection of negative energy is an operation that depends on the
frame where any measurement is to be performed. This suggests that
the sign of energy density for a quantum state may be a
coordinate-dependent quantity in quantum theory. \vspace{0.2cm}
\\{\bf Pacs}: 04.62. +v, 03.70. +k, 04.60. -m
\end{abstract}

\section{Introduction}
Although the  energy density of a field in classical physics  is
strictly positive, the local energy density in quantum field
theory  can be negative due to quantum coherence
effects\cite{EGJ}. The Casimir effect\cite{LMR} and squeezed
states of light\cite{WKHW} are two familiar examples which have
been studied experimentally. As a result, all the known pointwise
energy conditions in classical general relativity, such as the
weak energy condition and null energy condition, are allowed to be
violated. However, if the laws of quantum field theory place no
restrictions on negative energy, then it might be possible to
produce gross macroscopic effects such as violation of the second
law of thermodynamics\cite{L1,7}, traversable
wormholes\cite{10,11}, "warp drive"\cite{12}, and even time
machines\cite{11,13}. Therefore, a lot of effort has been made
toward determining the extent to which these violations of local
energy are permitted in quantum field theory. One powerful
approach is that of the quantum inequalities constraining the
magnitude  and duration of negative energy
regions\cite{L1,L2,L3,[10],[11],[12],[13]}.  Quantum inequalities
have been derived for scalar and electromagnetic fields in flat as
well as curved spacetimes\cite{PF97,Song,[11],FT99} and they have
also been examined in the background of evaporating black
holes\cite{BH1,Yu}. However, as far as the Dirac field is
concerned, not as much  work has been done. In this respect,
Vollick has shown that the superposition of two single particle
electron states can give rise to negative energy densities  and
demonstrated that the resulting energy densities obey quantum
inequalities which are derived for scalar and electromagnetic
fields\cite{DV1}. He has also given a quantum inequality for Dirac
fields in two-dimensional spacetimes\cite{DV2} using arguments
similar to those of Flanagan's\cite{EEF}. However, there does not
seem much hope of generalizing this argument beyond the two
dimensions. It is worth noting that the existence of quantum
inequalites for the Dirac (and Majorana) field in general
4-dimensional globally hyperbolic spacetimes was recently
established \cite{FV}.

In this paper, we will  examine the negative energy densities for
more general states that are the  superposition of two
multi-electron-positron states, and  discuss whether there are any
inequalities constraining the magnitude of negative energy when it
appears and its life time.  We will work in the units where
$c=\hbar=1$ and take the signature of the metric to be $(+ ~- ~-
~-).$

\section{Quantum states with negative energy densities}\label{sec2}

For the Dirac field  Lagrange density is
\begin{eqnarray}
\emph{L}=\frac{i}{2}\overline{\psi}\gamma^{\mu}\tensor\partial_{\mu}\psi-
m\overline{\psi}\psi.
\end{eqnarray}
The symmetrized stress tensor is given by
\begin{eqnarray}
T_{\mu\nu}=\frac{i}{4}\Big[\overline{\psi}\gamma^{\mu}\tensor\partial_{\nu}\psi+
\overline{\psi}\gamma^{\nu}\tensor\partial_{\mu}\psi\Big].
\end{eqnarray}
The field operator can be expanded as
\begin{eqnarray}
\psi(x)=\sum_{k}\sum_{\alpha=1,2}\Big[b_\alpha(k)u^\alpha(k)e^{ik\cdot
x}+ d_\alpha^\dagger(k)v^\alpha(k)e^{-ik\cdot x}\Big],
\end{eqnarray}
where the mode functions are taken to be
\begin{eqnarray}\label{eq:4}
u^\alpha(k)=\left(\begin{array}{l l}{\sqrt{\frac{\omega+m}{2\omega
V}} \phi^\alpha}\\{\frac{\bf{\sigma}\cdot\bf k}
{\sqrt{2\omega(\omega+m)V}}\phi^\alpha}\end{array}\right),
\end{eqnarray}
\begin{eqnarray}\label{eq:5}
v^\alpha(k)=\left(\begin{array}{l l}{\frac{\bf{\sigma}\cdot\bf
k}{\sqrt{2\omega(\omega+m)V}}\phi^\alpha}\\\sqrt{\frac{\omega+m}{2\omega
V}}\phi^\alpha\end{array}\right),
\end{eqnarray}
and $\phi^{1\dag}=(1,0)$, $ \phi^{2\dag}=(0,1)$. Here
$b_\alpha(k)$ and $b_\alpha^\dag(k)$ are the annihilation and
creation operators for the electron, respectively, while
$d_\alpha(k)$ and $d_\alpha^\dag(k)$ are the respective
annihilation and creation operators for the positron. The four
operators anticommute except in the cases $ \{b_\alpha(k),
b^{\dagger}_{\alpha'}(k')\}=\{d_\alpha(k),
d^{\dagger}_{\alpha'}(k')\}=\delta_{\alpha,\alpha'}\delta_{k,k'}.
$ The renormalized expectation value of the energy density, i.e.,
$\langle:T_{00}:\rangle$, in an arbitrary quantum state, is
\begin{eqnarray}\label{eq:7}
\langle\rho\rangle&=&\frac{1}{2}\sum_{k,k'}\sum_{\alpha,\alpha'}(\omega_k+\omega_{k'})\times\nonumber\\
&&\times[\langle b_\alpha^\dag(k) b_{\alpha'}(k')\rangle
u^{\dag\alpha}(k)u^{\alpha'}(k') e^{-i(k-k')\cdot x}
+\langle d_{\alpha'}^\dag(k')d_\alpha(k)\rangle v^{\dag\alpha}(k)v^{\alpha'}(k')e^{i(k-k')\cdot x}]\nonumber\\
&&+\frac{1}{2}\sum_{k,k'}\sum_{\alpha,\alpha'}(\omega_{k'}-\omega_k)\times\nonumber\\
&&\times[\langle d_\alpha(k)b_{\alpha'}(k')\rangle
v^{\dag\alpha}(k)u^{\alpha'}(k')e^{i(k+k')\cdot x} -\langle
b_\alpha^\dag(k)d_{\alpha'}^\dag(k')\rangle
u^{\dag\alpha}(k)v^{\alpha'}(k')e^{-i(k+k')\cdot x}]\;.
\end{eqnarray}

Now consider a state vector of the form
\begin{eqnarray} \label{eq:8}
|\Psi\rangle=\frac{1}{\sqrt{1+\lambda^{2}}}\big[|~a(q;j)~\rangle+\lambda|~b(l;n)~\rangle\big],
\end{eqnarray}
where $|~a(q;j)\rangle$ and $|~b(l;n)~\rangle$  are two
multi-particle states with the first symbol in the bracket
indicating the number of electrons and the second symbol the
number of  positrons. For example, we can write
$|~a(q;j)\rangle=|~k_1s_1,k_2s_2,\cdots,k_qs_q;\overline{k'_1s_1},\overline{k'_2s_2},\cdots,\overline{k'_js_j}~\rangle,$
and
$|~b(l;n)\rangle=|~k'_1s'_1,k'_2s'_2,\cdots,k'_ls'_l;\overline{k_1s'_1},\overline{k'_2s'_2},\cdots,\overline{k'_ns'_n}~\rangle\;.$
 Plugging Eq.~(\ref{eq:8}), Eq.~(\ref{eq:4}) and
Eq.~(\ref{eq:5}) into Eq.~(\ref{eq:7}),  we find
\begin{eqnarray}\label{eq:rho}
\langle\rho\rangle=\frac{1}{1+\lambda^{2}}\Bigg[\frac{1}{V}\Big(\sum_{r=1}^q\omega_{k_r}+\sum_{t=1}^j\omega_{\overline{k}_t}\Big)
+
(f_1+f_2+f_3+f_4)\lambda+\frac{1}{V}\Big(\sum_{f=1}^l\omega_{k_f}+\sum_{g=1}^n\omega_{\overline{k}_g}\Big)\lambda^{2}\Bigg]\;,
\end{eqnarray}
where
\begin{eqnarray}\label{eq:f1}
f_1&=&\frac{1}{2}\sum_{k,k'}\sum_{\alpha,\alpha'}(\omega_k+\omega_{k'})\times\nonumber\\
&&\times\Big[\langle~a(q,j)~|b_\alpha^\dag(k)
       b_{\alpha'}(k')|~b(l,n)~\rangle
       +\langle~b(l,n)~|b_\alpha^\dag(k)b_{\alpha'}(k')|~a(q,j)\rangle\Big]\times\nonumber\\
&&\quad \times u^{\dag\alpha}(k)u^{\alpha'}(k') e^{-i(k-k')\cdot
x}\;,
\end{eqnarray}
\begin{eqnarray}\label{eq:f2}
f_2&=&\frac{1}{2}\sum_{k,k'}\sum_{\alpha,\alpha'}(\omega_k+\omega_{k'})\times\nonumber\\
&&\times\Big[\langle~a(q,j)~|d_{\alpha'}^\dag(k')d_\alpha(k)|~b(l,n)~\rangle~
+\langle~b(l,n)~|d_{\alpha'}^\dag(k')d_\alpha(k)|~a(q,j)\rangle~\Big]\times\nonumber\\
&&\quad\times v^{\dag\alpha}(k)v^{\alpha'}(k')e^{i(k-k')\cdot
x}\;,
\end{eqnarray}
\begin{eqnarray}\label{eq:f3}
f_3=\frac{1}{2}\sum_{k,k'}\sum_{\alpha,\alpha'}(\omega_{k'}-\omega_k)&\Big[&\langle~a(q,j)~|d_\alpha(k)b_{\alpha'}(k')|~b(l,n)~\rangle
v^{\dag\alpha}(k)u^{\alpha'}(k')e^{i(k+k')\cdot
x}\nonumber\\
&&-\langle~b(l,n)~|b_\alpha^\dag(k)d_{\alpha'}^\dag(k')|~a(q,j)\rangle
u^{\dag\alpha}(k)v^{\alpha'}(k')e^{-i(k+k')\cdot x}\Big]\;,
\end{eqnarray}
and
\begin{eqnarray}\label{eq:f4}
f_4=\frac{1}{2}\sum_{k,k'}\sum_{\alpha,\alpha'}(\omega_{k'}-\omega_k)
&\Big[&\langle~b(l,n)~|d_\alpha(k)b_{\alpha'}(k')|~a(q,j)\rangle\Big]v^{\dag\alpha}(k)u^{\alpha'}(k')e^{i(k+k')\cdot
x}\nonumber\\
&&-\langle~a(q,j)~|b_\alpha^\dag(k)d_{\alpha'}^\dag(k')|~b(l,n)~\rangle
u^{\dag\alpha}(k)v^{\alpha'}(k')e^{-i(k+k')\cdot x}\Big]\;.
\end{eqnarray}
 Obviously, the first and the last term in Eq.(\ref{eq:rho})
 are always positive. Therefore, $\langle\rho\rangle$ can be
 negative only when the second term is non-vanishing.
 There are only four such cases.  {\it Case 1}.  The number of electrons and the number of positrons in
$|a\rangle$ are the same as those in  $|b\rangle$ respectively.
And there is only one different single electron state in these two
states. Here only $f_1$ is nonzero. {\it Case 2}.  The number of
electrons and the number of positrons in $|a\rangle$ are the same
as those in  $|b\rangle$ respectively. And there is only one
different single positron state in these two states. Here only
$f_2$ survives. {\it Case 3}. Two states are the same except for
that there is one more single electron state and one more single
positron state in  $|b\rangle$. Here only $f_3$ does not vanish. {
\it Case 4}. Two states are the same except for that there is one
more single electron state and one more single positron state in
$|a\rangle$. Here only $f_4$ is not equal to zero. Only in these
four cases can the energy density of the superposition state be
negative and all other possible cases all give rise to positive
results.
 Now we will discuss case 1 and case 3 in detail to see how negative energy can arise and if  certain
 quantum inequalities can be satisfied . It is easy to see that case 2 and case 4 are similar to case 1 and case 3 respectively.
\paragraph{case 1}
Let the two different single electron states in $|~a(q;j)\rangle$
and $|~b(l;n)\rangle$ to be characterized by
$(k_\epsilon,s_\epsilon)$ and $(k_\tau,s_\tau)$, respectively, and
for simplicity, take $k_\epsilon=k_{\epsilon_y}$,
$k_\tau=k_{\tau_z}$, $s_\epsilon=2$,$s_\tau=1.$
 Eq.~(\ref{eq:rho}) now reads
\begin{equation} \label{eq:9}
\langle\rho\rangle=\frac{1}{(1+\lambda^2)V}\bigg[\lambda^2\Big(E_0+\omega_{k_{\tau_z}}\Big)
+\lambda\beta_1 +\Big(E_0+\omega_{k_{\epsilon_y}}\Big) \bigg]\;,
\end{equation}
where
\begin{eqnarray}
E_0=\sum_{r=1}^{q-1}\omega_{k_r}+\sum_{t=1}^j\omega_{\overline{k}_t}\;,
\end{eqnarray}
\begin{equation}
\beta_1=\frac{k_{\epsilon_y}k_{\tau_z}
(\omega_{k_{\epsilon_y}}+\omega_{k_{\tau_z}})\sin\theta_1}
{2\sqrt{\omega_{k_{\epsilon_y}}\omega_{k_{\tau_z}}(\omega_{k_{\epsilon_y}}+m)
(\omega_{k_{\tau_z}}+m)}}\;,
\end{equation}
and  $\theta_1=(k_{\epsilon_y}-k_{\tau_z})\cdot x$.  $E_0$ is the
total energy of $q-1$ electrons and $j$ positrons. Note that the
energy density  $\langle\rho\rangle$ is that of a positive
constant part plus a part propagating at the speed of light in the
spacetime. Therefore, the sign of the energy could depend on the
location and time where any measurement is to be taken.  From
Eq.~(\ref{eq:9}) we know that $\langle\rho\rangle$ will be
negative if
\begin{eqnarray}\label{eq:cond1}
\beta_1^2>4(E_0+\omega_{k_{\tau_z}})(E_0+\omega_{k_{\epsilon_y}}),
\end{eqnarray}
 and if
\begin{eqnarray}
\label{eq:lamdaCond1} \frac{-\beta_1-\sqrt{\beta_1^2-
    4(E_0+\omega_{k_{\tau_z}})(E_0+\omega_{k_{\epsilon_y}})}}
    {2(E_0+\omega_{k_{\tau_z}})}<\lambda<\frac{-\beta_1+\sqrt{\beta_1^2-
    4(E_0+\omega_{k_{\tau_z}})(E_0+\omega_{k_{\epsilon_y}})}}
    {2(E_0+\omega_{k_{\tau_z}})}\;.
\end{eqnarray}
Let us now discuss if the quantum states could be manipulated to
satisfy Eq.~(\ref{eq:cond1}). In order to show that this is
possible, consider the ultrarelativistic limit,
$k_{\tau_z},k_{\epsilon_y} \gg m$. It then follows that
\begin{eqnarray}
\label{eq:16}
\beta_1=\frac{1}{2}(\omega_{k_{\tau_z}}+\omega_{k_{\epsilon_y}})\sin{\theta_1},
\end{eqnarray}
Substituting Eq.~(\ref{eq:16}) into  Eq.~(\ref{eq:cond1}), we have
\begin{eqnarray}\label{eq:18}
\sin^2\theta_1>\frac{16(E_0+\omega_{k_{\tau_z}})(E_0+\omega_{k_{\epsilon_y}})}
{(\omega_{k_{\tau_z}}+\omega_{k_{\epsilon_y}})^2}.
\end{eqnarray}
For Eq.~(\ref{eq:18}) to hold, it's necessary that
\begin{eqnarray}
\label{eq:cond2}
16(E_0+\omega_{k_{\tau_z}})(E_0+\omega_{k_{\epsilon_y}})
\leq(\omega_{k_{\tau_z}}+\omega_{k_{\epsilon_y}})^2\;.
\end{eqnarray}
 Eq.~(\ref{eq:cond2}) is satisfied if
\begin{eqnarray}
\omega_{k_{\epsilon_y}}\leq 7\omega_{k_{\tau_z}}+8E_0
  -4\sqrt{(3\omega_{k_{\tau_z}}+5E_0)(\omega_{k_{\tau_z}}+E_0)}\;,
\end{eqnarray}
or
\begin{eqnarray}
\omega_{k_{\epsilon_y}}\geq 7\omega_{k_{\tau_z}}+8E_0
  +4\sqrt{(3\omega_{k_{\tau_z}}+5E_0)(\omega_{k_{\tau_z}}+E_0)}\;.
\end{eqnarray}
Therefore, if the quantum states are manipulated in such a way
that the above conditions are met and $\lambda$ is chosen
according to Eq.~(\ref{eq:lamdaCond1}), then the energy density
for a quantum state of the form (\ref{eq:8}) can be made negative
at some places in space at some time.

\paragraph{case 3.}

Let the single positron and electron states in
 $|~b(l,n)\rangle$  that don't exist in $|~a(q,j)\rangle$ with be characterized by
 $(\overline{k}_\tau,\overline{s}_\tau)$ and
$(k_\varepsilon,s_\varepsilon)$ respectivel and further take $
k_\varepsilon=k_{\varepsilon_y}$,
$\overline{k}_\tau=\overline{k}_{\tau_z}$, $s_\varepsilon=1$ and
$\overline{s}_\tau=2$  as an example to study how  negative energy
density arises in this case. Now, the energy density becomes
\begin{equation} \label{eq:density3}
\langle\rho\rangle=\frac{1}{(1+\lambda^2)V}\Bigg[\lambda^2\bigg(E_a+\omega_{k_{\varepsilon_y}}+\omega_{\overline{k}_{\tau_z}}\bigg)+
\lambda\beta_3+E_a\Bigg]\;,
\end{equation}
where
\begin{equation}
E_a=\sum_{r=1}^q\omega_{k_r}+\sum_{t=1}^j\omega_{\overline{k}_t}\;,
\end{equation}
\begin{equation}
\beta_3=
\frac{(\omega_{\overline{k}_{\tau_z}}-\omega_{k_{\varepsilon_y}})k_{\varepsilon_y}}
{2\sqrt{\omega_{k_{\varepsilon_y}}(\omega_{k_{\varepsilon_y}}+m)}}
\sqrt{\frac{\omega_{\overline{k}_{\tau_z}}+m}{\omega_{\overline{k}_{\tau_z}}}}\sin\theta_3\;,
\end{equation}
and $\theta_3=(\overline{k}_{\tau_z}+k_{\varepsilon_y})\cdot x.$
Note that here again the energy density  $\langle\rho\rangle$ is
that of a positive constant part plus a part propagating at the
speed of light in the spacetime. It is easy to see that
$\langle\rho\rangle$ will be negative if
\begin{eqnarray}
\label{eq:condcase3}
\beta_3^2>4(E_a+\omega_{k_{\varepsilon_y}}+\omega_{\overline{k}_{\tau_z}})E_a\;
\end{eqnarray}
and
\begin{eqnarray}
\label{eq:lamdacond3}
 \frac{-\beta_3-\sqrt{\beta_3^2-4(E_a+\omega_{k_{\varepsilon_y}}+\omega_{\overline{k}_{\tau_z}})E_a}}
{2(E_a+\omega_{k_{\varepsilon_y}}+\omega_{\overline{k}_{\tau_z}})}<\lambda<\frac{-\beta_3+\sqrt{\beta_3^2-4(E_a+\omega_{k_{\varepsilon_y}}+\omega_{\overline{k}_{\tau_z}})E_a}}
{2(E_a+\omega_{k_{\varepsilon_y}}+\omega_{\overline{k}_{\tau_z}})}\;.
\end{eqnarray}
In the ultrarelativistic limit,
\begin{eqnarray}
\label{eq:beta3}
\beta_3=\frac{1}{2}(\omega_{\overline{k}_{\tau_z}}-\omega_{k_{\varepsilon_y}})
\sin\theta_3,
\end{eqnarray}
Substituting Eq.~(\ref{eq:beta3}) into Eq.~(\ref{eq:condcase3})
yields
\begin{eqnarray}
\sin^2\theta_3
>\frac{16(E_a+\omega_{k_{\varepsilon_y}}+\omega_{\overline{k}_{\tau_z}})E_a}
{(\omega_{\overline{k}_{\tau_z}}-\omega_{k_{\varepsilon_y}})^2}\;.
\end{eqnarray}
For the above inequality to admit a solution, we must require that
\begin{eqnarray}
16(E_a+\omega_{k_{\varepsilon_y}}+\omega_{\overline{k}_{\tau_z}})E_a
\leq(\omega_{\overline{k}_{\tau_z}}-\omega_{k_{\varepsilon_y}})^2\;.
\end{eqnarray}
And this is satisfied if
\begin{eqnarray}
\omega_{\overline{k}_{\tau_z}}\le
\omega_{k_{\varepsilon_y}}+8E_a\-
4\sqrt{2\omega_{k_{\varepsilon_y}}E_a+5E_a^2}\;,
\end{eqnarray}
or
\begin{eqnarray}
\omega_{\overline{k}_{\tau_z}}\ge
\omega_{k_{\varepsilon_y}}+8E_a\+
4\sqrt{2\omega_{k_{\varepsilon_y}}E_a+5E_a^2}\;.
\end{eqnarray}
Henceforth, if the quantum states are manipulated in such a way
that the above conditions are met and $\lambda$ is chosen
according to Eq.~(\ref{eq:lamdacond3}), then it is possible to
produce energy density  for a quantum state of the form
(\ref{eq:8}) at some places in space at some time.

It is interesting to note that the conditions derived above do not
apply when $\omega_{\overline{k}_{\tau_z}}=
\omega_{k_{\varepsilon_y}}$,  and when this happens $\beta_3$ is
zero, thus the energy density is positive. This reveals that in
the center of mass frame of the electron-positron pair in the
state $|~b(l,n)\rangle$, the local energy density for the
superposition state of the form (\ref{eq:8}) is always a positive
constant. Therefore for a given particle content of the state,
whether it is possible to detect negative energies is dependent
upon the frame in which any measurement is to be carried out. This
suggests that the sign of the energy density for a quantum state
may well be a coordinate-dependent quantity. It is worth noting
that the question of the observer dependence of negative energy
for scalar fields was also discussed in two dimensions \cite{BRF}.

\section{negative energy and quantum inequalities}
In the last Sect., we have found that under certain conditions,
the energy density of the superposition state of two
multi-particle states can be negative. Now, we want to demonstrate
that the larger the magnitude of this negative energy, the shorter
the duration that it persists.  For simplicity, we will consider
the ultrarelativistic limit with
$\omega_{k_{\tau_z}}\gg\omega_{k_{\epsilon_y}}$ and
$\omega_{k_{\tau_z}}\gg E_0$ for case 1 and
$\omega_{k_{\tau_z}}\gg E_a$ for case 3, then both
Eq.~(\ref{eq:9}) and  Eq.~(\ref{eq:density3}) become
\begin{equation} \label{eq:291}
\langle\rho\rangle=\frac{\lambda\omega_{k_{\tau_z}}}{(1+\lambda^2)V}\bigg[\lambda
+\frac{1}{2}\sin\omega_{k_{\tau_z}}(t-x) \bigg]
\end{equation}
and the condition for negative energy to arise is now
$-1/2<\lambda < 0$. Therefore, the energy density is that of a
constant positive background plus propagating wave at the speed of
light that alternates between negative and positive. At a fixed
spatial point, the total energy can dip to negative for a certain
period of time. The minimum value of $\langle\rho\rangle$ at a
fixed point $x$ is given by
\begin{equation} \label{eq:rhomin}
\langle\rho\rangle_{min}=\frac{\lambda\omega_{k_{\tau_z}}}{(1+\lambda^2)V}\Big(\lambda
+\frac{1}{2}\Big).
\end{equation}
At the same time, the length of time when the energy density is
negative is
\begin{eqnarray} \label{eq:duration}
\triangle t={ (\pi-2\sin^{-1}2|\lambda|)\over\omega_{k_{\tau_z}} }
={2\over\omega_{k_{\tau_z}}}\cos^{-1}(2|\lambda|)={2\phi\over
\omega_{k_{\tau_z}}}\;,
\end{eqnarray}
where $\phi\in (0,\pi). $ One can see that the larger the
magnitude of the negative energy $-\langle\rho\rangle_{min}V$ (or
equivalently the larger $\omega_{k_{\tau_z}}$ ), the shorter its
duration. In fact, we have
\begin{eqnarray} \label{eq:product}
V|\langle\rho\rangle_{min}|\triangle
t=-\frac{\lambda(2\lambda+1)}{(1+\lambda^2)}\phi\leq
-\frac{\lambda(2\lambda+1)\pi}{(1+\lambda^2)}=\pi g(\lambda)\;.
\end{eqnarray}
The function $g(\lambda)$ attains a maximum value of
$\sqrt{5}/2-1$, leading to that $\pi g(\lambda)\approx 0.37$.
Therefore, the negative energy satisfies the following quantum
inequality
\begin{eqnarray}
E\triangle t\le 1\;,
\end{eqnarray}
where we have defined that $E=V|\langle\rho\rangle_{min}|$. This
implies that the amount of negative energy that passes by a fixed
point in time $\Delta t$ is less than the quantum energy
uncertainty on that time scale, $\Delta t^{-1}$. It prevents
attempts of using quantum matter to produce bizarre macroscopic
effects. Finally, let us note that we can show, in essentially the
same way as in Ref.\cite{DV1}, that the sampled energy density for
the superposition states
\begin{eqnarray}\label{eq:SamEn}
\hat{\rho}=\frac{t_0}{\pi}\int_{-\infty}^{\infty}\frac{\langle\rho\rangle}{t^2+t_0^2}dt.
\end{eqnarray}
in the limits we considered above satisfies the quantum inequality
which was originally proven for scalar and electromagnetic fields.
\section{Conclusion}
We have examined the energy densities of quantum states that are
the superposition of two multi-electron-positron states. We have
found that the energy densities can be negative only when these
two states have the same number of electrons and positrons or when
one state has one more electron-positron pair than the other and
they are just positive constants for all the other possible cases.
In the cases in which negative energy could arise, we have shown
that the energy is that of a positive constant plus a propagating
part which oscillates between positive and negative, and if the
quantum states are properly manipulated, the energy can dip to
negative at some places for a certain period of time. It has been
demonstrated that the negative energy densities satisfy the
quantum inequality , which means that the product of its magnitude
and its duration is less than unity. Last but not the least, we
would like to note that in the case in which one state has one
more electron-positron pair, the energy density is a positive
constant in the center-mass frame of the pair in the state even it
can be negative in other frames. Therefore, for a given particle
content, the detection of negative energy is an operation that
depends on the frame where any measurement is to be performed.
This suggests that the sign of energy density  for a quantum state
may be a coordinate-dependent quantity in quantum theory.

\begin{acknowledgments}
We would like to acknowledge the support  by the National Science
Foundation of China  under Grant No. 10075019, the support by the
Natural Science Foundation of Hunan Province under Grant 02JJY2007
and the support by the Fund for Scholars Returning from Overseas
by the Ministry of Education of China.
\end{acknowledgments}


\end{document}